\documentclass[11pt,twoside]{article}
\usepackage{asp2010}

\resetcounters
\aspvolume{455}
\aspcpryear{2012}
\aspvoltitle{4$^{th}$ {\em Hinode} Science Meeting: Unsolved 
Problems and Recent Insights}
\aspvolauthor{L.~R.~Bellot Rubio, F.~Reale, and M.~Carlsson, eds.}

\begin{document}

\title{X-Ray Searches for Solar Axions}
\author{H.~S.~Hudson,$^{1,4}$ L.~W.~Acton,$^2$ E.~DeLuca,$^3$ I.~G.~Hannah,$^4$ K.~Reardon,$^5$ and K.~Van~Bibber$^6$}
\affil{$^1$Space Sciences Laboratory, UC Berkeley CA, USA 94720-7450}
\affil{$^2$Physics Department, Montana State Univ., PO Box 173500, Bozeman MT, USA 59717-0350}
\affil{$^3$Harvard-Smithsonian Center for Astrophysics, 60 Garden St, Cambridge MA, USA 02138}
\affil{$^4$School of Physics \& Astronomy, University of Glasgow, Glasgow, UK G12 8QQ}
\affil{$^5$INAF, Osservatorio Astrofisico di Arcetri, 50125 Firenze, Italia}
\affil{$^6$Naval Postgraduate School, Monterey CA, USA 93943; Lawrence Livermore National Laboratory, Livermore CA, USA 94550}

\begin{abstract}
Axions generated thermally in the solar core can convert nearly
directly to X-rays as they pass through the solar atmosphere via
interaction with the magnetic field.  The result of this conversion
process would be a diffuse centrally-concentrated source of few-keV
X-rays at disk center; it would have a known dimension, of order 10\%
of the solar diameter, and a spectral distribution resembling the
blackbody spectrum of the solar core.  Its spatial structure in detail
would depend on the distribution of mass and field in the solar
atmosphere.  The brightness of the source depends upon these factors
as well as the unknown coupling constant and the unknown mass of the
axion; this particle is hypothetical and no firm evidence for its
existence has been found yet.  We describe the solar magnetic
environment as an axion/photon converter and discuss the upper limits
obtained by existing and dedicated observations from three solar X-ray
observatories: \textit{Yohkoh}, RHESSI, and \textit{Hinode}.
\end{abstract}

\section{Introduction}
The axion is a hypothetical weakly-interacting particle whose
existence would solve certain problems in particle physics.  There are
also many implications for astrophysics, as described compendiously by
\cite{1996slfp.book.....R} and summarized more recently by
\cite{2007JPhA...40.6607R} and \cite{2009NJPh...11j5020Z}.  From the
solar perspective these implications are not trivial; a substantial
part of the solar core energy production (of order 0.1\%) could be
carried by these particles, if they existed
\citep{2009PhRvD..79j7301G}.  Because they are weakly interacting,
their emission by the core comprises an additional channel for solar
energy loss, as do the neutrinos.  Axions are created in the solar
core via the Primakoff effect \citep{1951PhRv...81..899P}, and this
process also allows the axions to convert back into photons via
coupling with an ambient magnetic field.  In this interaction the
resultant photon nearly preserves the incident momentum and energy of
the axion.  The solar axions mostly result from thermal photons in the
core (X-rays) interacting with a nuclear field as a virtual photon,
the escaping axion spectrum resembles the core blackbody distribution.


Many searches have been carried out for the solar axions, which are
attractive observationally because the source is relatively nearby and
the fluxes are (hypothetically) large.  These searches individually
cover different regions of the parameter space (the unknown axion
mass, and its coupling constant $g_{a\gamma}$).  These searches
include ground-based ``helioscopes'' or ``axion telescopes'', of which
the largest and most sensitive has been the \textit{CERN Axion Solar
Telescope} \citep[CAST; ][]{2007JCAP...04..010A}.  The ground-based
searches rely upon powerful magnets to convert the axions; the CERN
magnet has a field of about 9~T over a 10-m length.

In this paper we follow up on a pioneering search by 
\citet[][see also \citealt{2009NJPh...11j5020Z}]{1996PhLB..365..193C} 
for solar axions via solar X-ray observations.  This search was
carried out with the \textit{Yohkoh} Soft X-ray Telescope \citep[SXT;
][]{1991SoPh..136...37T}. Such a search makes use of the natural
magnetism of the solar atmosphere as a converter.  We extend this
search with \textit{Yohkoh} and describe also searches with RHESSI
\citep{2007ApJ...659L..77H,2010ApJ...724..487H} and with
\textit{Hinode/XRT} (Section~\ref{sec:searches}).  In such searches
one is at the mercy of the vagaries of the solar magnetic field for
the conversion.  The X-rays can propagate to near-Earth space where a
solar X-ray telescope can detect them, provided that the conversion
occurs high enough in the solar atmosphere for the photons to escape
without being first absorbed.  As described in
Section~\ref{sec:axion_basics}, the search is sensitive to an axion
mass range that depends upon density distribution in the atmosphere.
The run density and field in the quiet solar atmosphere are not
understood very precisely at present, and this is even truer of
different solar features such as faculae that may have much stronger
magnetic fields.  We discuss the issues involved in this knowledge in
Section~\ref{sec:b}.  A successful detection of an unambiguous axion
signal would be possible even knowing these properties, since the
axion X-ray signatures are so specific, but it would not be very
precise.  Indeed solar physicists would immediately want to make use
of this signature to study the solar magnetic field itself.

\section{Solar Magnetism}
\label{sec:b}
The solar magnetic field is extremely complicated
\citep[e.g.,][]{1993PhDT.......225H}.  Near the photosphere it is
neither a simple dipole, as eclipse pictures during solar minimum
suggest, nor is it random.  We observe the solar magnetic field mainly
via use of the Zeeman effect in photospheric spectral lines.  It
appears to consist of two independent components: the quiet Sun, in
which the field is highly correlated with the visible convective
motions in the photosphere, and the active regions.  Sunspots come and
go with the Hale cycle of alternating polarities, and the magnetic
flux that first appears in the active regions as spots appears to
diffuse away across the disk, ultimately concentrating in the polar
regions on the solar-cycle time scale.  This component of the field
thus has large-scale ordering both in space and time.  The
active-region fields may be orders of magnitude greater than those of
the quiet Sun.  The density structure of the solar atmosphere depends
on the magnetic field in a dynamic manner, and there is a high degree
of variability on small scales.  From this it is clear that to predict
the axion conversion rate in detail is impossible, and we can only
hope to estimate it statistically.  Indeed, for the quiet Sun, it may
be that MHD simulations \citep[e.g.,][]{2007ApJ...665.1469A} may
provide the simplest approach to this problem.

In the meanwhile there are several possible approaches.
\cite{1996PhLB..365..193C} simply assumed a dipole field 
consistent with the then-current understanding of the quiet-Sun field
structure.  The situation has changed dramatically in recent years,
with the discovery of strong horizontal fields in the lower solar
atmosphere \citep{2007ApJ...659L..77H,2008ApJ...672.1237L}.  These
fields have horizontal intensities of order 50~G, rather than the
vertical intensities an order of magnitude weaker that were commonly
assumed a decade ago \citep[e.g.,][]{2000asqu.book.....C}.  Indeed,
the small-scale field may be more intense still
\citep{2004Natur.430..326T}.  This makes a drastic difference for the
expected quiet-Sun flux owing to the $B^2$ conversion efficiency.
Unfortunately, we have little understanding of the vertical structure
of the small-scale fields.

In active regions the photospheric fields are easier to measure, and
much stronger, but again the extension of their structure is not well
understood.  The active-region corona has a low plasma beta (below
10$^{-3}$ for typical parameters), with relatively weak currents
appearing to thread it, and so a potential-field approximation makes a
plausible beginning \citep[see][for further
discussion]{2008ApJ...675.1637S}.  But the distribution of matter
within this field is complicated and there is no compelling theory
that describes, for example, the coronal pressure as a function of
position in an active region.

\section{Axion Conversion}
\label{sec:axion_basics}
\begin{figure}
\centering
\includegraphics[width=0.52\textwidth]{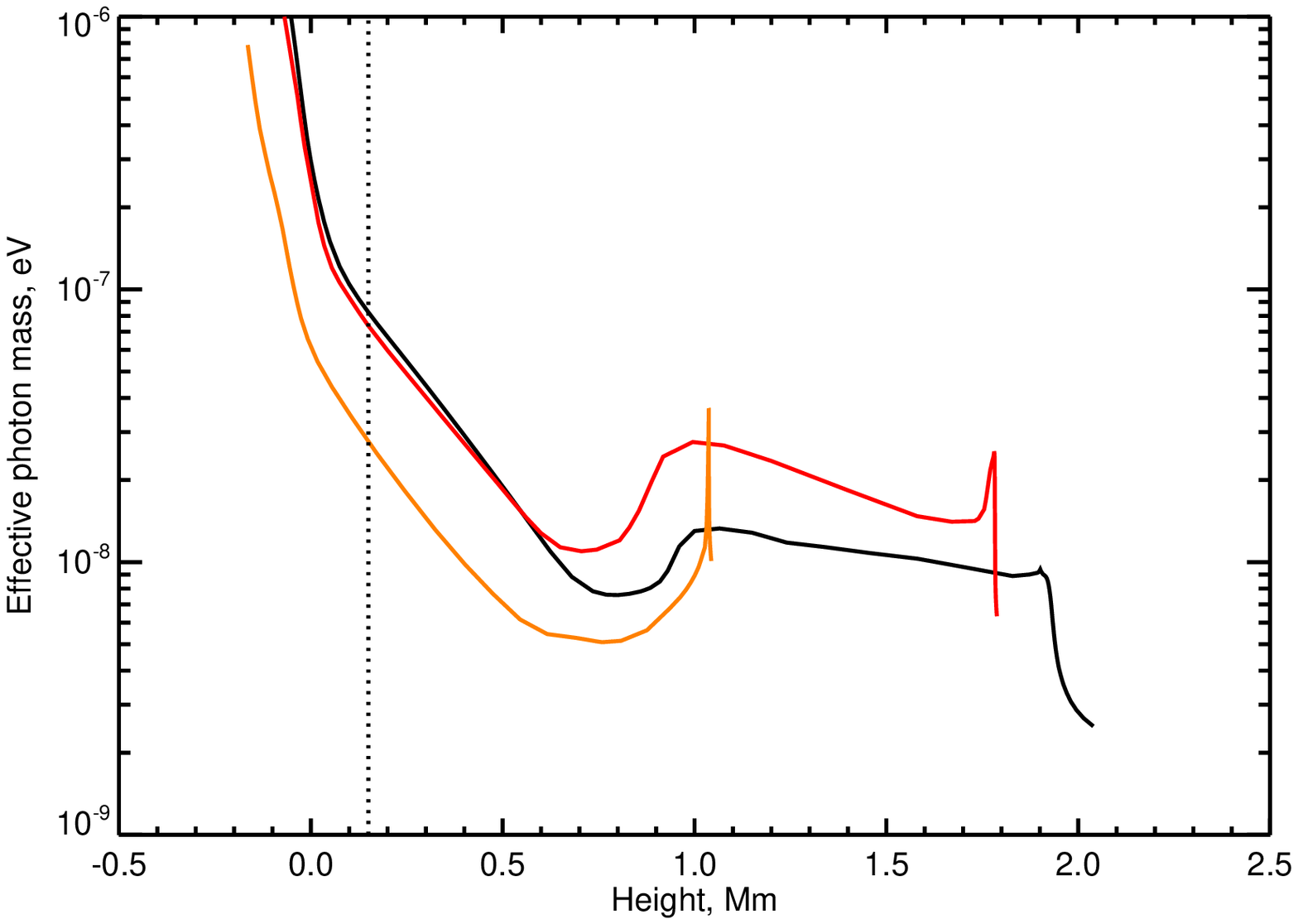}
\includegraphics[width=0.46\textwidth]{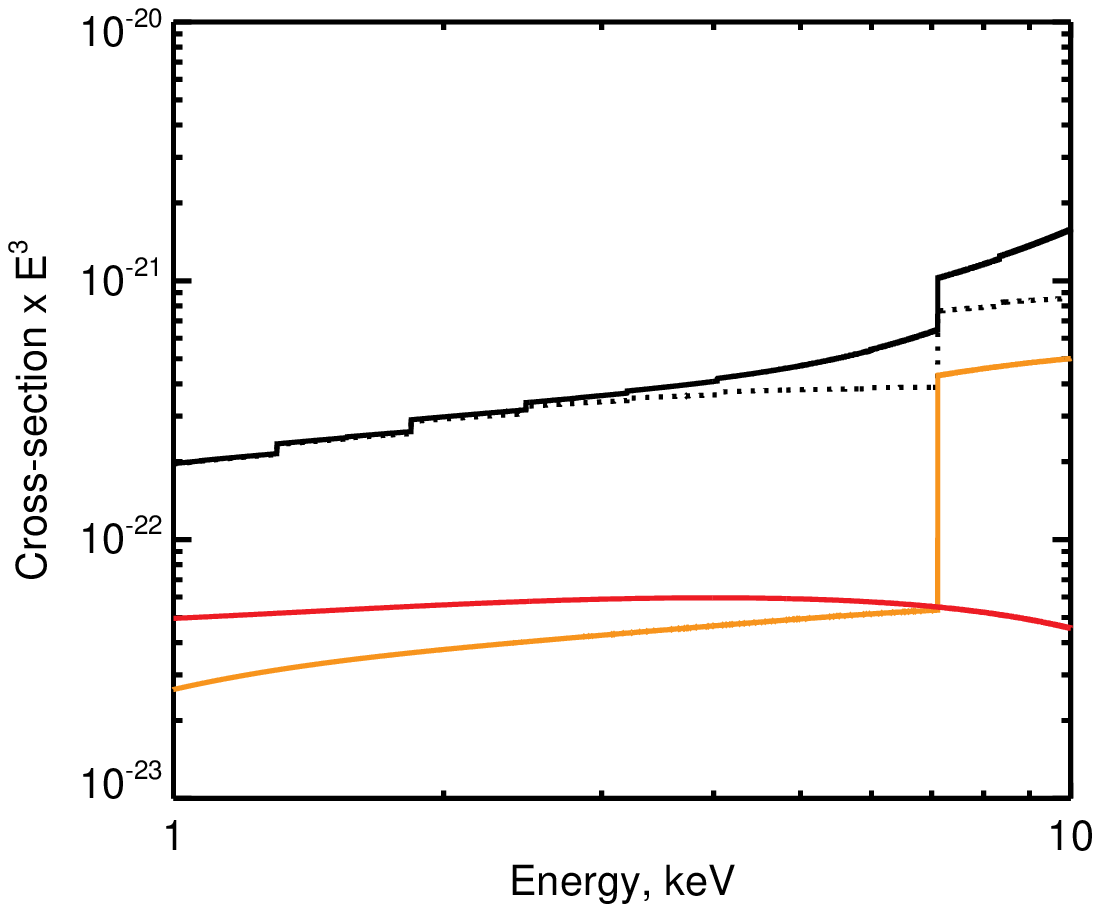}
\caption{Parameters of the solar atmosphere. \textit{Left:} 
effective photon mass in three standard solar model atmospheres from
\cite{2009ApJ...707..482F}---in order of increasing transition-region
height---sunspot umbra, facular region, and the quiet Sun.  The dotted
line shows the $\tau = 1$ height for 4.2~keV. \textit{Right:} the
absorption cross-section for X-rays in the quiet-Sun model.  The
photoeffect dominates in the 1-10~keV region; solid line shows the
total and the lower lines He and Fe (with its K-edge); the dotted line
shows the Thomson-scattering contributions.  These cross-sections use
\cite{2009ARA&A..47..481A} abundances in the quiet-Sun model and have
been multiplied by $E^3$ for clarity. }
\label{fig:2up}
\end{figure}

The X-ray opacity of the solar atmosphere at relevant energies largely
results from the photoeffect, as shown in Fig.~\ref{fig:2up}
(right), and $\tau = 1$ occurs low in the atmosphere.  This
corresponds to a relatively high density, depending upon the
particular solar feature and its dynamics.  The traditional approach
to estimating this and other parameters is through standard
``semi-empirical'' modeling, which aims to reproduce the emergent
spectrum with non-LTE radiative transfer but otherwise limited
physics; such models are 1D and time-stationary.  From Equation~6 of
\cite{2009NJPh...11j5020Z} we derive
\begin{equation}\label{eq:g}
{\rm d}P(\varepsilon, h) = g_{{\rm a}\gamma}^2 
\frac{B_\perp(h)^2}{q(h,m_{\rm a},\varepsilon)^2+\kappa(\varepsilon)^2/4} 
\, q(h,m_{\rm a},\varepsilon) e^{-\tau(\varepsilon)}\ {\rm d}l
\label{eq:cont}
\end{equation}
for the source function of the emergent X-ray photons, where
$\kappa(\varepsilon)$ is the mass absorption coefficient.  Here $q(h)
= (m_{\rm a}^2 - m_\gamma^2)/2E_{\rm a}$, with $m_\gamma$ the effective mass of
the photon in natural units (Fig.~\ref{fig:2up}, left), and
$\tau(\varepsilon) = \int_h^\infty{\kappa\ {\rm d}h}$.  
\citeauthor{2009NJPh...11j5020Z} also quote an approximate axion flux of
\begin{displaymath}
J_{\rm a} = {\rm d}\Phi_{\rm a}/{\rm d}E_{\rm a} = g_{10}^2\ 3.821 \times 10^{10} \, 
{\rm cm}^{-2} \, {\rm s}^{-1} \, {\rm keV}^{-1} (E_{\rm a}/{\rm keV})^3 / 
(e^{E_{\rm a}/1.103 \,{\rm keV}} -1),
\end{displaymath}
where $g_{10} = g_{{\rm a}\gamma}/10^{10}\ \mathrm{GeV}$.  The
observable X-ray flux $J_\varepsilon = \int_0^\infty{J_{\rm a} {\rm
d}P(\varepsilon,h) {\rm d}h} \propto g_{10}^4$ because of the two
conversions needed.

\begin{figure}
\centering
\includegraphics[width=0.49\textwidth]{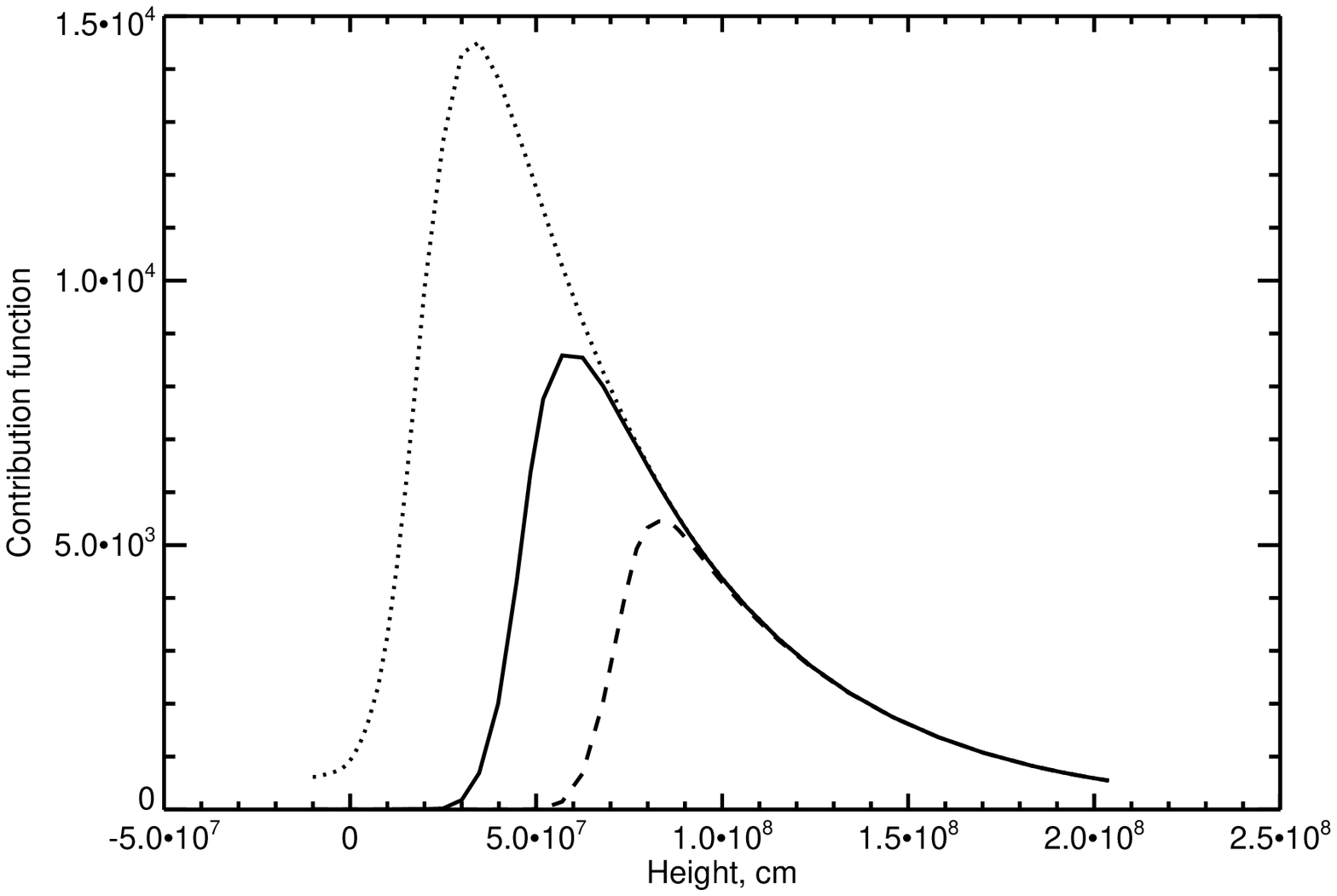}
\includegraphics[width=0.49\textwidth]{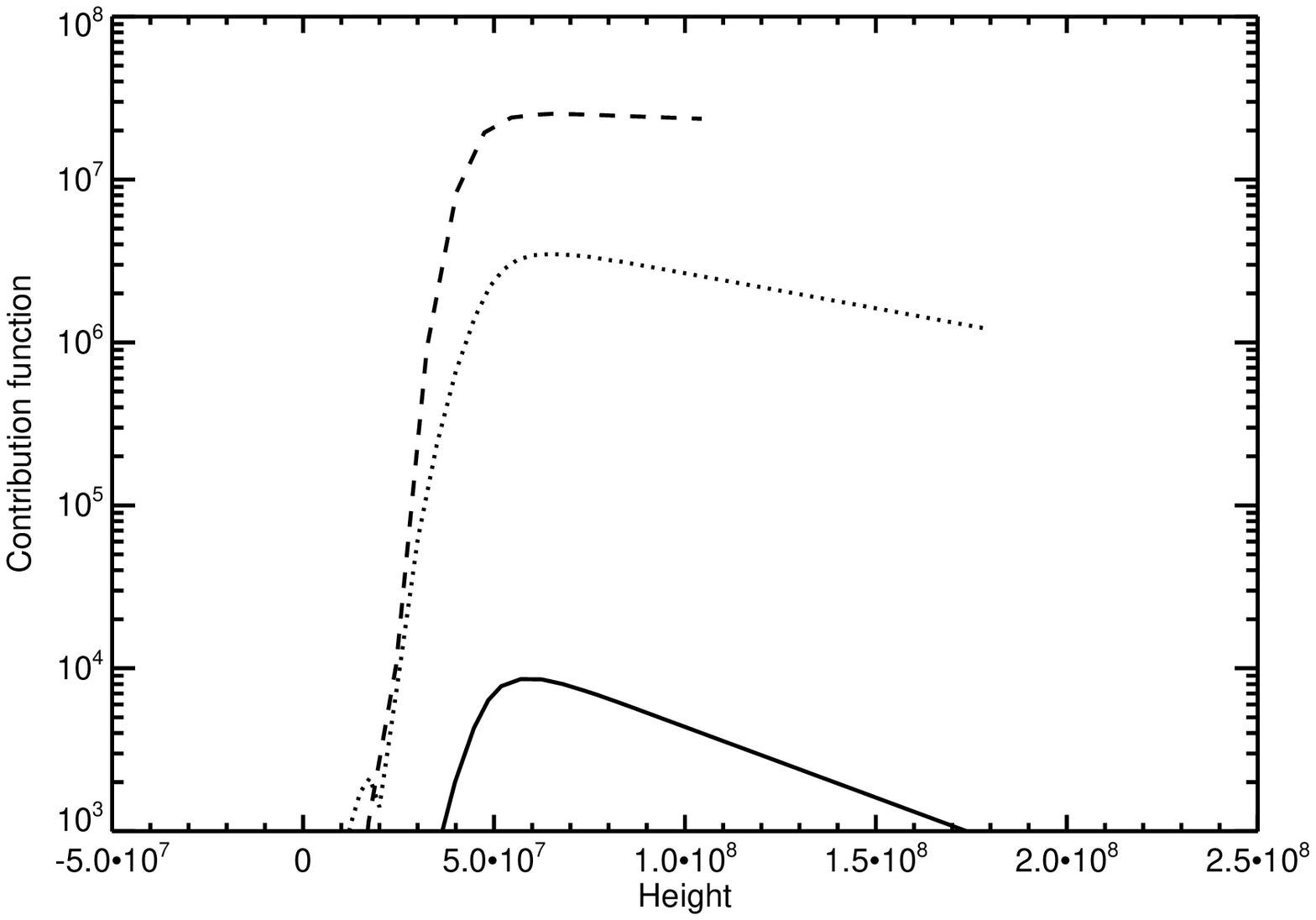}
\caption{\textit{Left:} the contribution function in the quiet-Sun 
model (solid), illustrating \textit{ad hoc} 10$\times$ variations 
upward (dashed) and downward (dotted) in density relative to the quiet-Sun model.
\textit{Right:} contribution functions from the three models 
(dashed, umbra; dotted, facula; solid, quiet Sun).}
\label{fig:2moreup}
\end{figure}

Axion conversion in the natural magnetic field of the Sun differs from
that in controlled laboratory conditions.  The values of $B_\perp$ and
density, and hence $m_\gamma$, vary in an ill-known manner through the
solar atmosphere.  For the purposes of this paper we have approximated
the magnetic field above the photosphere as a simple exponential with
photospheric magnitudes of [100, 1500, 3000]~G and scale heights of
[1, 2, 10]~Mm for three Fontenla (2009) semi-empirical models
describing quiet Sun, facula, and sunspot umbra respectively.
Figure~\ref{fig:2moreup} (left) shows contribution functions
calculated for these models and for a range of assumed densities via
Equation~\ref{eq:cont}, and Fig.~\ref{fig:2moreup} (right) shows the
three models considered.  Both panels assume an axion mass of $1\
\mu$eV, and the variation with density mainly reflects the
photoelectric absorption of the X-rays produced deep in the
atmosphere.

\section{The Searches}
\label{sec:searches}

The solar axion source, modulo the complexities discussed in
Sect.~\ref{sec:b}, should be an easy matter for X-ray astronomers.
The source has well-defined signatures in space, energy, and time.
Unfortunately the telescopes that we have for solar X-ray observation
are not as sensitive as those used for non-solar X-ray astronomy, and
there is competing emission from ordinary solar magnetic activity.
Because of the latter problem, the most sensitive searches (those
described below) are from sunspot minimum periods.  In addition to the
competition from solar activity (flares, microflares, X-ray bright
points, and hot active regions), there is also a slowly-varying corona.
In the absence of activity this background solar X-ray emission is so
low that it cannot easily be measured
\citep{1966JGR....71.5778P,2007ApJ...659L..77H,2010ApJ...724..487H},
and we do not know how bright the Sun is at the peak energy of the axion
component, 4.2~keV \citep[see][for a discussion of low-level solar
emission]{2008MNRAS.385..719C}.

The three solar X-ray telescopes used in the axion searches each have
different properties, as described in the following sections.
Generally the sensitivity of a given observation scales as the figure
of merit \citep{1975ARA&A..13..423P} given by
\begin{equation}
{\rm FOM} = \sqrt{A \Delta t \varepsilon \Delta E / B},
\label{eq:fom}
\end{equation}
where A is the collecting area (cm$^2$), $\Delta t$ the integration
time (s), $\varepsilon$ the detection efficiency, $\Delta E$ the
spectral band (keV), and $B$ the background counting rate in counts
cm$^{-2}$ s$^{-1}$ keV$^{-1}$.  We do not estimate the detection
efficiency parameter $\varepsilon$, which refers to the quantum
efficiency as well as other factors, such as the convolution of the
telescope spectral response with the known spectral distribution of
the axion source or the duty cycle of the observations.
Table~\ref{tab:fom} summarizes some of the parameters for the search
efforts discussed below.  Note that the information in this table is
inadequate, owing to limited space, for accurately estimating the
figure of merit via Equation~\ref{eq:fom}.

\begin{table}[!t]
\centering
\caption{Properties of solar axion searches}
\smallskip
\begin{tabular}{l c c c c c c}
\hline
Search & $\bar{E}$ (keV) & Area (cm$^2$) & $\Delta t$ (s) & $\Delta E$ (keV) & Target \\
\hline
\textit{Yohkoh} & 2.1 & 0.78 & $2 \times 10^5$ & 1.0 & Quiet Sun\\
RHESSI offpoint & 4.5 & 1.1 & $1.03 \times 10^6$ & 3.0 & Quiet Sun\\
RHESSI direct & 4.5 & 30 & --- & 3.0 & Spots \\
GOES sunspots & 4.2 & 2.3 & --- & 3.5 & Spots \\
\textit{Hinode}/energy & 1.53 & 2 & & 0.7 & Quiet Sun\\
\textit{Hinode}/histogram & 1.53 & 2 & & 0.7 & Quiet Sun\\
\hline
\label{tab:fom}
\end{tabular}
\end{table}

\subsection{Simple Photometric Searches (the ``Sunspot Flash'')}
\label{sec:flash}

Sunspot fields may attain many thousands of Gauss and have scales of
tens of Mm, and so active regions are a good place to search for
axionic X-rays \citep{1996PhLB..365..193C}.  In principle as a sunspot
group crosses disk center, it will become anomalously bright if axions
are converting because of the $B^2$ dependence; the competition from
ordinary forms of solar activity will also be intense but generally
concentrated in lower photon energies.  A search for a ``sunspot
flash'' near disk center would preferably be done in hard X-rays and
for an older sunspot region with relatively weak magnetic activity.
We note that the umbral field tends to be vertical, which does not
favor conversion, but that the field rapidly diverges and so large
volumes of large $B_\perp$ will be available at times during the
disk-center passage.

The standard GOES photometry, routinely available for decades, is
biased towards short wavelengths in its 0.5--4~\AA~band
\citep[e.g.,][]{2005SoPh..227..231W}.  Thus it (as well as RHESSI) has
a favorable spectral response.  The spectrum of axion X-rays,
integrated over the two GOES spectral bands, yields a hardness ratio
of about~2; by contrast the usual optically-thin thermal spectra due
to solar plasma activity have hardness ratios of order 0.001--0.01,
depending upon the plasma temperature.  As a zeroth-order check for
axion presence, we have examined the two-channel GOES spectral ratio
(nominally 0.5--4~\AA~and 1--8~\AA) for several strong active regions
and not seen any evidence of an axion-related flash at disk center.
Such a search could also readily be carried out with RHESSI or SphinX
\citep{2008JApA...29..339S} data and illustrates the use of the
\textit{spectral} and \textit{temporal} signatures; the imaging
instruments described below also employ the \textit{spatial} signature
to discriminate between axion-conversion and ordinary solar X-rays.

\subsection{The \textit{Yohkoh} Search}
\begin{figure}
\centering
\includegraphics[width=0.57\textwidth]{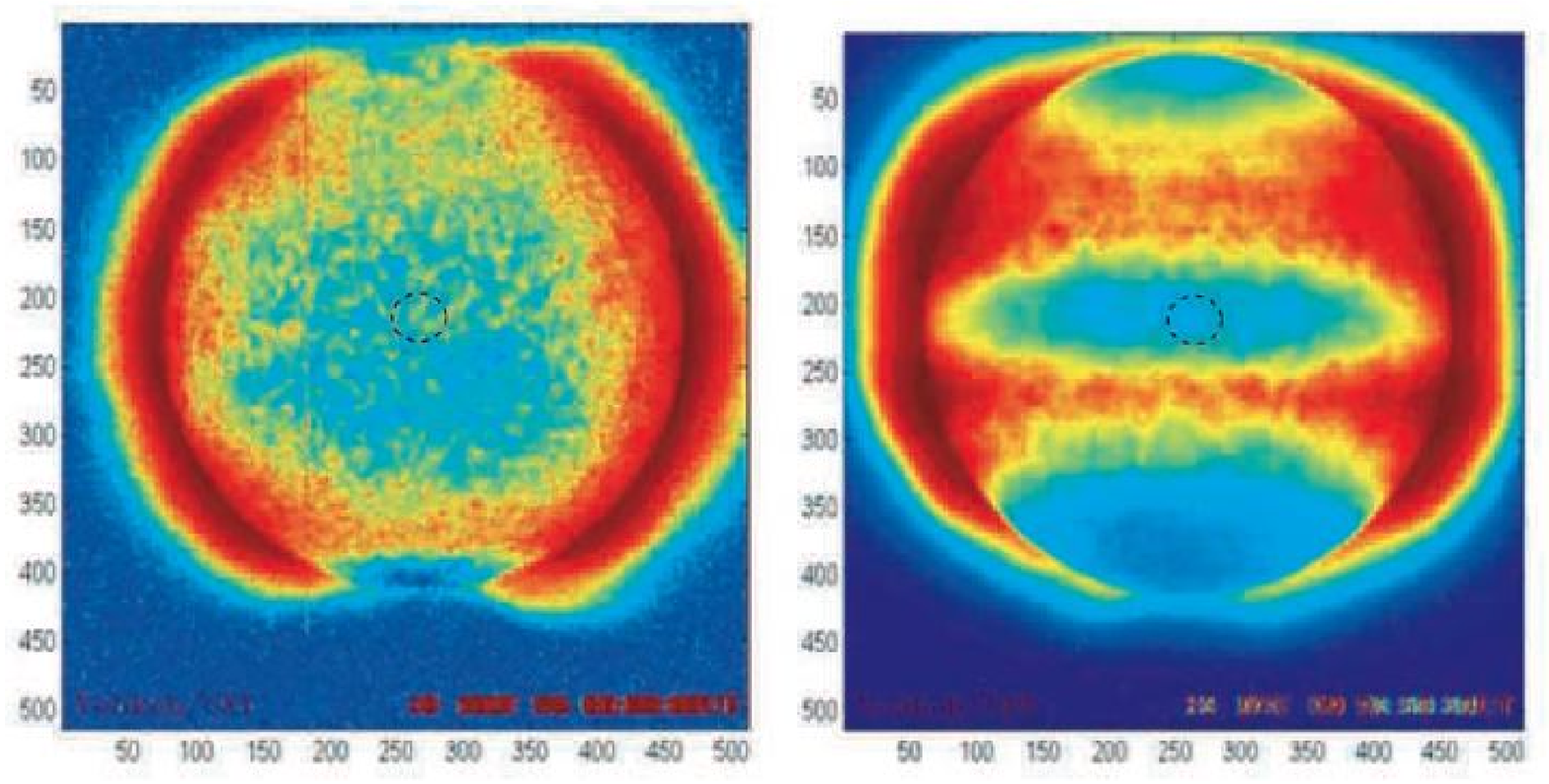}
\includegraphics[width=0.41\textwidth]{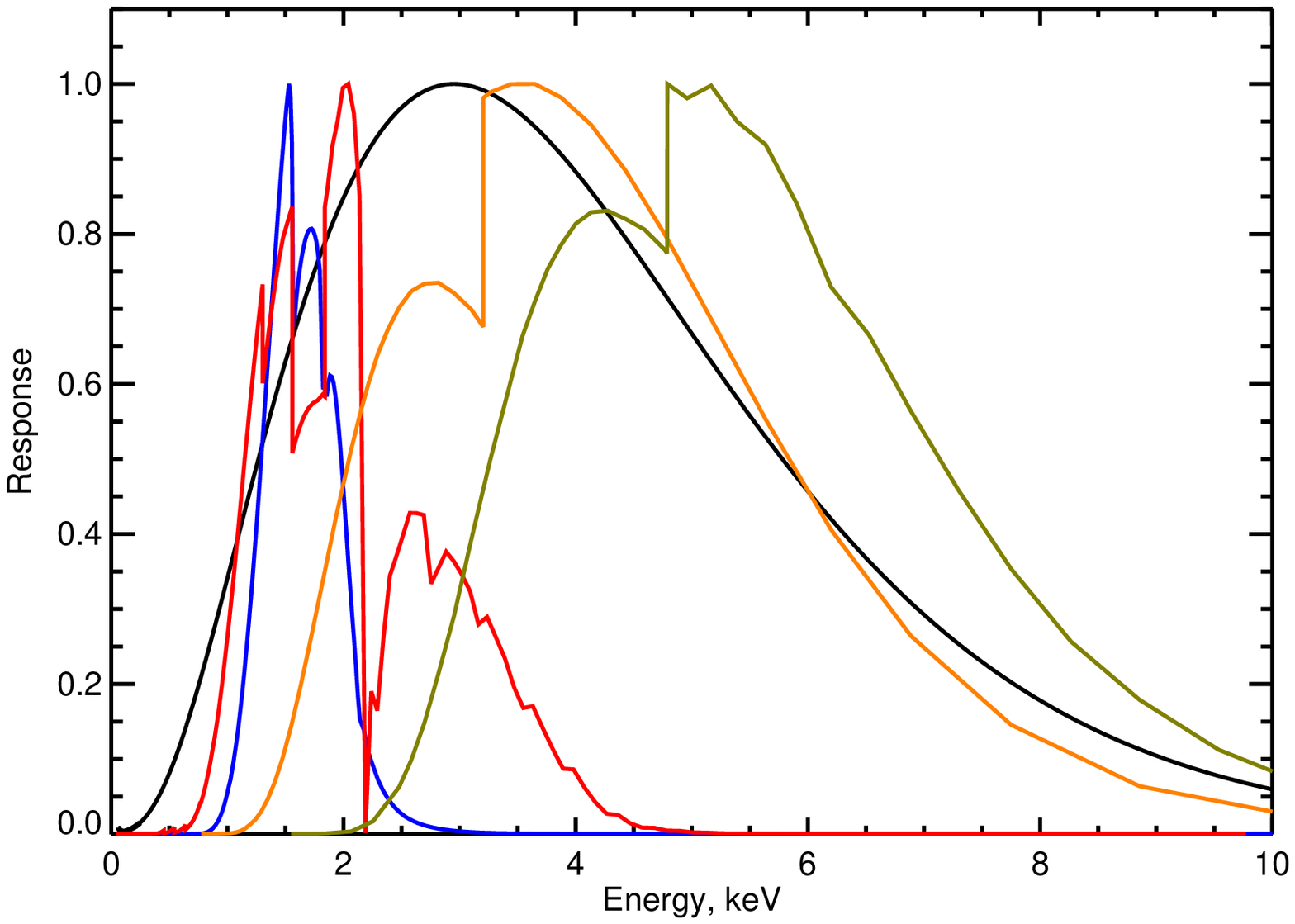}
\caption{\textit{Left:} quiet Sun, in a \textit{Yohkoh}/SXT image search for axion-related soft X-rays. \textit{Middle:} similar image sum for active times.
The dotted circle at Sun center shows the location of the expected
source \citep[from][]{2009NJPh...11j5020Z}.
\textit{Right:} spectral passbands for the various searches---blue, 
the \textit{Hinode}/XRT Be/Med filter; red, the \textit{Yohkoh}/SXT
AlMg filter; orange and olive, the GOES 1--8~\AA~and
0.4--4~\AA~passbands.  All of these are normalized to their peak values
and are weighted against the theoretical spectral distribution of the
axions (black).  }
\label{fig:zioutas}
\end{figure}

The \textit{Yohkoh} SXT operated with a two-reflection mirror system
at grazing incidence, and accumulated image charge on a
1024~$\times$~1024-pixel CCD via shuttered exposures through metallic
analysis filters to isolate given spectral ranges.  In terms of the
factors in Equation~\ref{eq:fom}, the effective area of this
instrument was about 0.78~cm$^2$ over a spectral bandwidth, depending
on the filter chosen, of about 1~keV, effectively at a peak energy in
the 1--2~keV range.  Figure~\ref{fig:zioutas} shows the morphology of
summed SXT soft X-ray images \citep{2009NJPh...11j5020Z}.  The image
on the left sums quiet periods from 1996, and the image on the right
shows more active times.  This view of the data is rather qualitative,
but shows no evidence for excess emission due to axions.

The quantitative analysis of the image data is relatively simple.  We
take the most appropriate exposures, sum them up, and set limits on a
source with the expected angular scale at disk center by differencing
against the image external to this region; the time-series
fluctuations in the pixel sums lead to an uncertainty and an estimate
of the upper limit.  We have obtained roughly 2~ksec of exposures in
the AlMg filter \citep{1991SoPh..136...37T} during the quietest times
in 1996, and from these images obtain an upper limit on the flux of
0.2~ph (cm$^2$~s~keV)$^{-1}$ at a mean energy 2.1~keV weighted against
the photon spectrum of the axion source.

\subsection{The RHESSI Search}
The RHESSI instrument differs conceptually from a traditional soft
X-ray telescope in that it employs modulation optics.
\cite{2007RScI...78b4501H} describes the methods used for 
RHESSI quiet-Sun observations.  This kind of optics permits
observations to high energies, including $\gamma$~rays, but has a high
background rate because there is no focusing.  Furthermore in the
standard observing mode the net A$\Omega$ product for the extended
axion source is reduced by the modulations necessary for image
formation.  The special observations devised for quiet-Sun viewing
\citep{2007RScI...78b4501H} require pointing RHESSI about 1$^\circ$
away from Sun center. \cite{2007ApJ...659L..77H,2010ApJ...724..487H}
have published the completed analysis for RHESSI observations taken in
this mode in the recent solar minimum (2005--2009).

RHESSI searches for axion X-rays could also be made via its direct modulation in data from its low-resolution subcollimators.
This would have the advantage of much longer integration times, since it would make use of essentially all of the data; in addition the data could be used to detect the ``sunspot flashes'' described in Section~\ref{sec:flash}. 
Neither of these approaches has been followed yet.

\subsection{The \textit{Hinode} Search}

The \textit{Hinode}/XRT instrument resembles SXT on \textit{Yohkoh} in
that it is a grazing-incidence soft X-ray telescope with CCD readout
in energy mode.  The lack of a mirror surface coating restricts XRT to
a mean photon energy of about 1.53~keV, as weighted by the theoretical
spectrum of the axions (Figure~\ref{fig:zioutas}, right panel).
During the recent solar minimum period XRT obtained almost 15\,000
exposures in the ``medium beryllium'' filter, at 11.6~s exposure time
in $8 \times 8$-pixel binning mode.  These image parameters optimize
transmission and suppression of readout noise in the camera.  The net
result is a total of more than 150~ksec exposure; this data set has
the best sensitivity for solar axion detection by XRT.

The choice of filter and the low count rates allow us to attempt a
novel spectral analysis of the data, namely the use of the pixel
histograms to do pulse-height spectroscopy
\citep[e.g.,][]{2007SoPh..240..387L} and thereby to decrease the 
background rate.  This is possible even though XRT works in energy
mode rather than as a photon counter.  In this approach a sequence of
images yields a signal histogram for each pixel.  Read noise and dark
noise populate the bottom range of the histogram, as do the
$\sim$0.1--0.2~keV photons due to ordinary solar activity.  The most
common signal level, for these data, is determined by the read noise
of the CCD.  A single true photoelectron at the mean energy expected
(determined by the convolution of the telescope efficiency vs. energy
and the axionic X-ray spectrum) would have a larger signal; ideally
this single-photoelectron response would produce an identifiable peak
in the histogram.  The $\sim$1.5~keV photons due to axion conversion
would appear well above this noise level, and the observed rate of
counts in these region of the histogram will have a background rate
limited mainly by ordinary non-solar background sources such as
Compton-scattered $\gamma$-rays, neutron-induced background, etc 
\citep[see, e.g.,][]{1975ARA&A..13..423P}.

\section{Discussion}

The observations we have described have not yet positively identified
any signature of solar axion emission, but considerable further work
in refining these searches is possible.  Table~\ref{tab:lim} assembles
representative values of the current upper limits derived from the
various.  The entries in the table are not definitive, except for the
RHESSI ``offpoint'' entry, for which the observational work is
essentially complete.  As noted in Sect.~\ref{sec:b}, there are large
unknowns in the physical parameters the conversion depends upon,
specifically upon the structure of the solar magnetic field.  To estimate
the mass range we have taken the simplest case, the quiet Sun, and
adopted reasonable guesses regarding the necessary physical
parameters.  These are that the horizontal field is 100~G at the
photosphere and falls off exponentially with a scale height of 1~Mm,
roughly the granulation scale.  The assumed density structure for the
quiet Sun follows Model 1001 of \cite{2009ApJ...707..482F}.

\begin{table}[!t]
\tabcolsep 1em
\centering
\caption{Current X-ray limits (two sigma)}
\smallskip
\begin{tabular}{l c c c c}
\hline
\noalign{\smallskip}
Search & $\bar{E}$ &   Flux limit &  g$_{10}$ limit\\
       &  (keV)   & (ph~cm$^{-2}$~s$^{-1}$~keV$^{-1}$) & ($\mu$eV) \\
\hline
GOES/long & 3.6 &  --- & ---\\
GOES/short & 4.8 &  --- & --- \\
\textit{Yohkoh}/AlMg & 2.04 &   1.2 & 0.4 \\
RHESSI offpoint& 5 &    340 & 1.8 \\
RHESSI direct & 5 &    --- &  --- \\
\textit{Hinode}/energy & 1.53 &   0.1 & 0.3 \\
\textit{Hinode}/histogram& 1.53 &   0.01 & 0.2 \\
\hline
\end{tabular}
\label{tab:lim}
\end{table}

\begin{figure}[!t]
\centering
\includegraphics[width=1.0\textwidth]{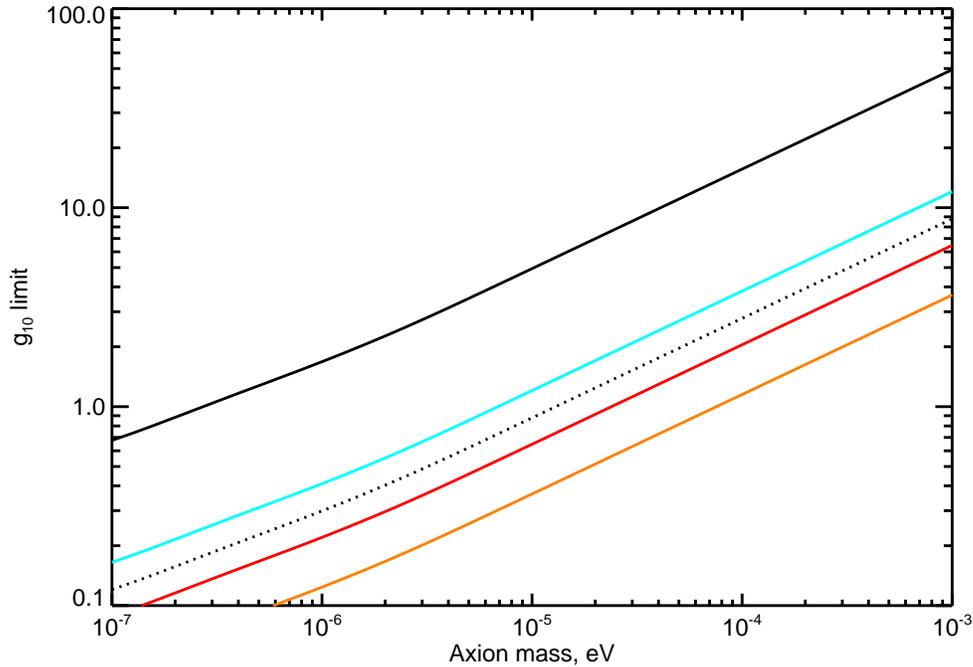}
\caption{Comparison of limits on $g_{10}$ from the three X-ray 
telescopes used for quiet-Sun integrations (RHESSI, black; SXT, blue;
XRT, red and gold for direct and histogram methods). The dotted line
shows the point of equality with the diffuse-component albedo as
estimated by \cite{2008MNRAS.385..719C}.  }
\label{fig:glim}
\end{figure}

Figure~\ref{fig:glim} shows the initial results from the three X-ray
telescopes.  The dotted line in the figure shows the limit that would
be obtained for an axion signal equal to the estimated solar albedo
due to cosmic X-rays \citep{2008MNRAS.385..719C}.  This is a practical
but not fundamental limit, since the spectral and spatial signatures
of the axion signal would still be available for deeper explorations.

\section{Conclusions}

This paper is a progress report on deep searches for solar X-rays
originating in axion emission from the core of the Sun, as converted
in the magnetic field in the solar atmosphere by the Primakoff effect.
None of the searches to date have revealed a definite signal, but the
limits on the coupling constant for the Primakoff effect can rival
those obtained with laboratory techniques for axion masses below about
$10^{-4}$~$\mu$eV.  The main emphasis here has been to describe the
solar framework for these searches, rather than to give definitive
upper limits.  The limits, indeed, will be quite uncertain for some
time given our lack of knowledge of the physical conditions in the
solar atmosphere.

There are several lines of research that could improve on the present
limits, notably the use of any of the three data sets available in
searches for the sunspot flash signature \citep[][see
Section~\ref{sec:flash}]{1996PhLB..365..193C}.  This kind of
observation depends on the fortuitous occurrence of a large quiescent
sunspot passing close to disk center, not a common occurrence because
of the brevity (about one day) of the transit.  Such searches could be
carried out in many other databases of non-imaging solar X-ray data,
preferably ones in which a useful spectral signature could be
incorporated with the temporal signature due to the disk passage of
sunspots.  Here a temporal signature could also be applied to the
search even below the (soft) limit imposed by the diffuse component.
Another advantage of the sunspot fields is their extension into the
corona, a volume not considered in the calculation of contribution
functions given in this paper, and which will give correspondingly
more sensitivity in the search.

\acknowledgments
This research was supported by NASA under contract NAS 5-98033 for
RHESSI.  We thank many persons on the \textit{Yohkoh}, RHESSI, and
\textit{Hinode} teams, and especially the latter two because of the
special observations obtained for these searches.

\bibliography{hinode4}

\end{document}